\documentclass[final,3p,linenumbers,compress]{elsarticle} 

\usepackage{lineno}
\usepackage{hyperref}
\usepackage{amssymb}
\usepackage{mathtools}
\usepackage{soul}

\usepackage{footnote}

\usepackage{amsmath,color,soul}

\begin{document}

\begin{frontmatter}

\title{Search for the Chiral Magnetic Effect in Au+Au collisions at $\sqrt{s_{_{\rm{NN}}}}=27$ GeV with the STAR forward Event Plane Detectors}

\author{
B.~E.~Aboona$^{53}$,
J.~Adam$^{15}$,
L.~Adamczyk$^{2}$,
J.~R.~Adams$^{38}$,
I.~Aggarwal$^{40}$,
M.~M.~Aggarwal$^{40}$,
Z.~Ahammed$^{59}$,
D.~M.~Anderson$^{53}$,
E.~C.~Aschenauer$^{6}$,
J.~Atchison$^{1}$
% ...
}

\address{\rm{(STAR Collaboration)}}

\address{$^{1}$Abilene Christian University, Abilene, Texas   79699}
\address{$^{2}$AGH University of Science and Technology, FPACS, Cracow 30-059, Poland}
% ...
\address{$^{6}$Brookhaven National Laboratory, Upton, New York 11973}
% ...
\address{$^{15}$Czech Technical University in Prague, FNSPE, Prague 115 19, Czech Republic}
% ...
\address{$^{38}$The Ohio State University, Columbus, Ohio 43210}
% ...
\address{$^{40}$Panjab University, Chandigarh 160014, India}
% ...
\address{$^{53}$Texas A\&M University, College Station, Texas 77843}
% ...
\address{$^{59}$Variable Energy Cyclotron Centre, Kolkata 700064, India}
% ...

\begin{abstract}
A decisive experimental test of the Chiral Magnetic Effect (CME) is considered one of the major scientific goals at the Relativistic Heavy-Ion Collider (RHIC) towards understanding the nontrivial topological fluctuations of the Quantum Chromodynamics vacuum. In heavy-ion collisions, the CME is expected to result in a charge separation phenomenon across the reaction plane, whose strength could be strongly energy dependent. The previous CME searches have been focused on top RHIC energy collisions.  
In this Letter, we present a low energy search for the CME in Au+Au collisions at $\sqrt{s_{_{\rm{NN}}}}=27$ GeV. 
We measure elliptic flow scaled charge-dependent correlators relative to the event planes that are defined at both mid-rapidity $|\eta|<1.0$ and at forward rapidity $2.1 < |\eta|<5.1$. 
We compare the results based on the directed flow plane ($\Psi_1$) at forward rapidity and the elliptic flow plane ($\Psi_2$) at both central and forward rapidity. The CME scenario is expected to result in a larger correlation relative to $\Psi_1$ than to $\Psi_2$, while a flow driven background scenario would lead to a consistent result for both event planes. In 10-50\% centrality, results using three different event planes are found to be consistent within experimental uncertainties, suggesting a flow driven background scenario dominating the measurement. We obtain an upper limit on the deviation from a flow driven background scenario at the 95\% confidence level. This work opens up a possible road map towards future CME search with the high statistics data from the RHIC Beam Energy Scan Phase-II. 
\end{abstract}


\end{frontmatter}

\section{Introduction}
Relativistic heavy-ion collisions are the ideal testing ground for the theory of strong interaction and its symmetries. The hot and dense medium produced in these collisions has been conjectured to be accompanied by an axial charge asymmetry, where the parity ($P$) and charge-parity ($CP$) are violated locally, leading to a difference in number of right-handed and left-handed quarks~\cite{Kharzeev:1999cz,Kharzeev:2004ey,Kharzeev:2007tn,Kharzeev:2007jp,Fukushima:2008xe}. 
Such an imbalance can result in a separation of electric charge in the direction of the extremely strong (10$^{14}$ T) magnetic field (B), produced by the protons in the colliding heavy-ions~\cite{Kharzeev:2007jp,Skokov:2009qp}. 
This phenomenon is known as the Chiral Magnetic Effect (CME).
Observations consistent with the CME have been reported in condensed matter systems~\cite{Li:2014bha}. However, their verification in relativistic collision-produced medium is still pending.

In heavy-ion collisions, the CME is expected to cause a charge separation across the reaction plane determined by the impact parameter and the beam direction. This is because the reaction plane is correlated to the direction of the magnetic field. Therefore, the CME will lead to preferential emission of positively and negatively charged particles into opposite sides of the reaction plane~\cite{Voloshin:2004vk,Ajitanand:2010rc}.
Finding conclusive experimental evidence for this phenomenon has become one of the major scientific goals of the heavy-ion physics program at the Relativistic heavy-ion Collider (RHIC)~\cite{Abelev:2009ac,Abelev:2009ad,Abelev:2012pa,Adamczyk:2013hsi, Adamczyk:2013kcb, Adamczyk:2014mzf, Adam:2015vje,STAR:2019xzd} during the past decade. 
Possible signals for this effect have also been extensively studied at the Large Hadron Collider (LHC)~\cite{Khachatryan:2016got,Acharya:2017fau,Sirunyan:2017quh}.
However, measurements sensitive to CME are also sensitive to background correlations~\cite{Schlichting:2010qia,Bzdak:2010fd,Wang:2009kd} and the two sources are very difficult to separate.
Therefore, recent experimental measurements have focused on disentangling the signal and background~\cite{Kharzeev:2015znc,Tribedy:2017hwn,Zhao:2018skm,Zhao:2019hta,Li:2020dwr}, providing upper limits on the observability of the CME~\cite{Acharya:2017fau,Sirunyan:2017quh} or providing data-driven baselines for background estimates~\cite{STAR:2019xzd,Khachatryan:2016got,STAR:2020gky}. 

The measurements at the LHC have provided upper limits on the observability of the CME in 2.76 TeV and 5.02 TeV Pb+Pb collisions ~\cite{Acharya:2017fau, Sirunyan:2017quh}. 
Two recent measurements from STAR (Solenoidal Tracker at RHIC) have provided upper limits on the CME fraction in Au+Au collisions at $\sqrt{s_{_{\rm NN}}}=200$ GeV. 
The first one used the pair invariant mass dependence of the CME sensitive charge separation observable $\Delta \gamma$ and found an upper limit of CME signal to be 15\% of the inclusive result at the 95\% confidence level (CL)~\cite{STAR:2020gky}. %
The second one exploited the difference of the CME sensitive observables and elliptic flow as the main background source with respect to the spectator neutron plane and participant plane. Such analysis found a hint of positive signal in mid-central events with 1-3$\sigma$ significance~\cite{STAR:2021pwb}. 
Among extensive experimental efforts in disentangling signal and background, the most controlled and precise measurement has been done in collisions of isobars $^{96}_{44}$Ru+$^{96}_{44}$Ru and $^{96}_{40}$Zr+$^{96}_{40}$Zr at the top RHIC energy~\cite{STAR:2021mii}. 
Under the standards of a blind analysis with a set of predefined criteria, no evidence consistent with a signal for the CME was found in isobar collisions~\footnote{Two- and three-particle non-flow contribution to the CME measurement by spectator and participant planes were studied in Ref.~\cite{Feng:2021pgf} as well as to incorporate the multiplicity difference between the two isobars that can modify the baseline for a CME scenario.}. 

An outstanding question is the behavior at lower collision energy. The change of collision energy affects the prerequisites for the CME such as the magnetic field lifetime, the domain size of axial charge imbalance, and the presence of a medium where quarks and gluons are deconfined and the chiral symmetry of Quantum Chromodynamics (QCD) is restored~\cite{McLerran:2013hla,Kharzeev:2020jxw,Ikeda:2020agk,Cartwright:2021maz,Ghosh:2021naw}. 
Furthermore, the background contributions to the CME are also expected to change with the collision energy. 
Despite the theoretical progress, a quantitative prediction for the collision energy dependence of the CME signal remains challenging~\cite{Jiang:2016wve,Shi:2017cpu}. 
Therefore, a dedicated effort on the CME search at collision energies below $\sqrt{s_{_{\rm NN}}}=200$ GeV is very desirable and timely. The first low energy CME search from STAR under the Beam Energy Scan program Phase-I (BES-I) was reported in Ref.~\cite{Adamczyk:2014mzf}. An important observation was that by lowering collision energies the charge separation decreases and eventually disappears at $\sqrt{s_{_{\rm NN}}}=7.7$ GeV. Such an observation might be driven by the disappearance of either signal or background sources of charge separation. Further investigation of CME driven charge separation at lower energies have been limited by statistics of BES-I data and poor resolution of event plane determination at lower energies. Several previous flow measurements from STAR indicate that a partonic phase, necessary for the CME phenomenon, may be created in Au+Au collisions above $\sqrt{s_{_{\rm NN}}}>10$ GeV~\cite{STAR:2013ayu,STAR:2014clz}. This gives us the necessary impetus for CME search above $\sqrt{s_{_{\rm NN}}}>10$ GeV with improved detector capabilities.

In this letter, we present an analysis of a high statistics data sample of Au+Au collisions at $\sqrt{s_{_{\rm NN}}}=27$ GeV collisions taken by the STAR detector in the year of 2018 with the newly
installed highly-segmented Event Plane Detectors (EPDs)~\cite{Adams:2019fpo}. The EPD is one of the major upgrades added to the STAR detector for the Beam Energy Scan phase II (BES-II) program. It covers the pseudorapidty window of $2.1 < |\eta| < 5.1$ symmetrically around the mid-rapidity and significantly improves the event plane resolution at forward rapidity. We would like to note that EPDs increase the resolution of the reconstracted first order event plane by a factor two  compared to the previously used Beam Beam Counter (BBCs) that had much coarser granularity~\cite{Adams:2019fpo}. We measure elliptic flow scaled charge dependent correlations relative to event planes using the EPDs and the Time Projection Chamber (TPC) ~\cite{Anderson2003659} at mid-rapidity $|\eta|<1.0$. Then we compare the results using the directed flow plane ($\Psi_1$) at forward rapidity and the elliptic flow planes ($\Psi_2$) at both central and forward rapidity. The $\Psi_1$ plane determined by the EPDs is dominated by the large directed flow of protons and has stronger correlation to the magnetic field direction than $\Psi_2$ plane does. As a result, the CME scenario is expected to yield in a larger charge separation across $\Psi_1$ than that of $\Psi_2$, while a flow driven background scenario would lead to a consistent result for both the event planes. We search for evidence of the CME driven charge separation and provide an upper limit on deviations from a flow driven background scenario. 

We have organized this paper as follows. In Sec.~\ref{sec:detector}, we introduce the detectors and data sample followed by the analysis techniques in Sec.~\ref{sec:analy_tech}. We discuss the systematic uncertainty sources in Sec.~\ref{sec:syserr}. We present the results in Sec.~\ref{sec:result} and a summary in Sec.~\ref{sec:summary}.

\section{Detectors and data sample}
\label{sec:detector}
STAR was the only operational detector at RHIC during the collection of Au+Au 27 GeV data in the year of 2018. The main subsystems of STAR used for this analysis are the TPC, Time-of-Flight (ToF) detector~\cite{Llope:2003ti}, Vertex Position Detectors (VPDs)~\cite{Llope:2014nva}, and the EPDs. 
Charged particles are detected within the range of $|\eta|\!<\!1$, over full $2\pi$ azimuthal coverage and transverse momentum ($p_T$) larger than $0.2$ GeV/$c$ using the STAR TPC situated inside a 0.5 T solenoidal magnetic field. For this $p_T$ range we estimate the tracking efficiency of the TPC to range from $77\%$ to $86\%$ using embedding simulations based on the \textsc{geant} ~\cite{Fine:2000gn}. 
The TPC is used to reconstruct the position of the primary vertices of collisions along the beam direction ($V_{z}$) and along the radial direction transverse to the beam axis ($V_r$).
For the current analysis we restrict the positions of primary vertices within $|V_{z}|<40$ cm and $V_{r}<2$ cm. To reduce the contamination from  secondary charged particles, we only select tracks with a distance of closest approach (DCA) to the primary vertex of less than 3 cm. We also require at least fifteen ionization points in the TPC for selecting good tracks.
STAR collected minimum-bias events by requiring the coincidence of signals from the Zero Degree Calorimeters (ZDCs)~\cite{Adler:2000bd}, on either side of the interaction region, at the rate of 0.5-2 kHz. Among these minimum bias events we identified approximately $0.023\%$ out-of-time pile-up of two events that we remove by studying the correlation between the number of TPC tracks and the number of tracks matched with a hit in the ToF detector. We also require good events have at least one TPC track matched to the ToF. After these event cuts, approximately 300 million minimum bias events become available for our analysis.

The EPD system used for event plane measurements consists of two wheels located $\pm 3.75$ m away from the center of the TPC, covering approximately $2.1 <|\eta| < 5.1$ in pseudorapidity and $2\pi$ in azimuth. Each wheel consists of 12 “supersectors”, each of which consists of 31 plastic scintillator tiles. Each tile is connected to a silicon photomultiplier (SiPM) via optical fiber. Charged particles emitted in the forward and backward directions produce a signal distribution in the EPD tiles with identifiable peaks corresponding to 1, 2, 3, $\cdots$ minimally ionizing particles (MIPs). A threshold value of 0.3 MIP is used as a default parameter for hit identification. We use the MIP weighted hit distribution to reconstruct the event planes in our analysis. Details of the EPDs can be found in Ref.~\cite{Adams:2019fpo}.
\begin{figure}[htb]
    \centering
    \includegraphics[width=0.45\textwidth]{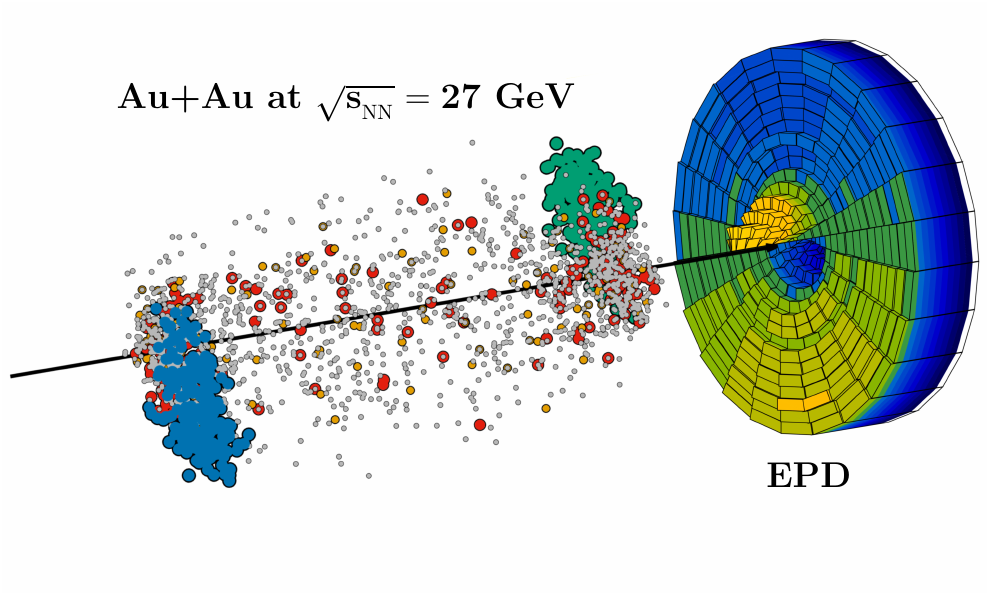}
    \caption{Cartoon to demonstrate the EPD detector acceptance and response to directed flow from both spectator protons and participant particles. The left shows the sum of ten simulated UrQMD~\cite{Bass:1998ca} events with identical event planes and the right (on a different scale) represents the response of the EPDs to real data (yellow representing more counts, blue representing fewer) with approximately matching event plane. Beam rapidity for these 27 GeV events is $Y_{\rm beam}$ = 3.4 which falls within the acceptance of the EPD $(2.1 < |\eta| < 5.1)$. Forward spectator protons are represented by the green color in the UrQMD cartoon and by the matching detector hits near the center of the EPD (yellow peak-like structure). Produced particles, colored grey (pions), yellow (kaons) and red (protons) in the UrQMD cartoon, are responsible for the peak near the outer edge of the EPD opposite to the inner peak in azimuthal angle. At this energy, the inner EPD sectors detect beam fragments, stopped and spectator protons which have the opposite sign of directed flow compared to the forward produced particles that are detected by the outer EPD sectors.}
    \label{fig:epd_acceptance}
\end{figure}

In 27 GeV Au+Au collisions, a unique capability can be achieved with this detector as illustrated in Fig.~\ref{fig:epd_acceptance}. In this figure we show the positions of different particles from ten simulated UrQMD~\cite{Bass:1998ca} events with identical event planes. In addition, we also show the response of the EPDs to incident particles from many events using real data with approximately matching event plane.  
The rapidity of beam remnants and other breakup products from the colliding beam (Y$_{\mathrm{beam}}$ =3.4) falls in the acceptance of the EPDs (2.1 $< |\eta| <$ 5.1). Therefore, the EPDs can measure the directed flow ($v_1$) at forward rapidity due to the beam fragments and stopped protons. 
Interestingly, the directed flow changes sign between the inner half ($|\eta| > {\rm Y}_{\rm beam}$) and outer half ($|\eta| < {\rm Y}_{\rm beam}$) of the EPDs. Observation of large forward directed flow at $|\eta| > {\rm Y}_{\rm beam}$ and sign change at ${\rm Y}_{\rm beam}$ were made from PHOBOS measurements~\cite{Back:2005pc}. The EPDs were built to measure the $\Psi_1$ plane corresponding to such a large forward directed flow that is expected to be a proxy for the reaction plane, particularly for $|\eta| > {\rm Y}_{\rm beam}$~\footnote{ 
Note that $\Psi_1$($|\eta| > Y_{\rm beam}$) measured from the EPDs is not the plane of only spectators and should not be confused and compared to the commonly used spectator neutron planes measured using the ZDC~\cite{Adler:2001fq,SN0448}.}. 

The benefit of using the EPDs for CME search is that we expect the forward $\Psi_1$($|\eta|$ $>$ Y$_{\rm beam}$) plane to be more correlated to the magnetic field than the elliptic flow plane determined by particles from mid and forward rapidities. This is supported by our UrQMD simulations~\cite{Bass:1998ca}: 1) $\Psi_1$($|\eta|$ $>$ Y$_{\rm beam} $) is dominated by charged hadrons, most of which ($70\%$) are protons that are fragments and spectators and not produced in collisions, and 2){ we find that the correlation of B-field with $\Psi_1(|\eta|>$Y$_{\rm beam} )$ is $18.9 \pm 1.2\%$ and $11.1\pm 2.0\%$ stronger compared to the same with $\Psi_2$ from TPC ($\Psi_2(|\eta|<1)$) and EPD($\Psi_2(|\eta|<$Y$_{\rm beam}))$, respectively when the strength is estimated by the quantity $\gamma_{B}(\Psi_n)=\left<\cos(2\Psi_B-2\Psi_n)\right>$.}~\footnote{ We estimate the magnitude $B$ and the direction ($\Psi_B$) of the B-field at the central point of the participant zone defined by the average weighted positions of the participants for that event. We then estimate various harmonic event planes such as $\Psi_2(|\eta|<1$), $\Psi_2(|\eta|<{\rm Y_{beam}})$ and $\Psi_1(|\eta|>{\rm Y_{beam}}$) similar to what has been used in the data analysis, using the final state particles produced in the same UrQMD event. 
}.

\section{Analysis techniques}
\label{sec:analy_tech}
The primary CME sensitive charge separation observable~\footnote{Along with the conventional $\gamma-$correlator the novel R-variable and the signed balance functions have been proposed as alternative observables for CME search. Recently, members of the STAR collaboration has also performed model calculations to  demonstrate the sensitivity of different CME  observables~\cite{Choudhury:2021jwd}. We do not explore such observable in this study and only stick to the studies of $\gamma$-correlator.}
, the  $\gamma$-correlator, is defined as:
\begin{linenomath}
\begin{equation}
  \gamma(\phi^{\alpha},\phi^{\beta}) =\langle\cos{(\phi^{\alpha}+\phi^{\beta}-2\Psi_{\rm RP})}\rangle,  
\end{equation}
\end{linenomath}
where $\phi^{\alpha}$ and $\phi^{\beta}$ denote the azimuthal angles ($\phi$) of charge particles, and $\Psi_{\rm RP}$ is the reaction plane angle~\cite{Voloshin:2004vk}. The charge separation is quantified by the difference between the $\gamma$-correlators measured for the opposite-sign (OS) and the same-sign (SS) particles defined as, 
\begin{linenomath}
\begin{equation}
  \Delta\gamma = \gamma_{OS} - \gamma_{SS}.
\end{equation}
\end{linenomath}
In addition, we introduce the scaled charge separation correlator:
\begin{linenomath}
\begin{equation}
    \Delta\gamma/v_2\,\,\, {,\ \rm where} \,\,\, v_2=\left<\cos(2\phi-2\Psi_{\rm RP})\right>. 
\end{equation}
\end{linenomath}
The normalized quantity $\Delta\gamma/v_2$ is better to account for the trivial scaling expected from a flow driven background due to resonance decay and local charge conservation~\cite{Voloshin:2004vk,Schlichting:2010qia}.
As a proxy for the $\Psi_{\rm RP}$, we use the first order event plane, $\Psi_{1,\  \mathrm{Y_{beam}}<|\eta|<5.1}$, from the directed flow of forward protons.
We compare such measurements with the charge separation across: 1) the second-order plane driven by elliptic flow of the forward participants, i.e., using $\Psi_{\rm RP}=\Psi_{2,\ 2.1<|\eta|<\mathrm{Y_{beam}}}$, and 2) the second-order plane of produced particles at mid-rapidity, i.e., with $\Psi_{\rm RP}=\Psi_{2,\ |\eta|<1}$. If the background is entirely due to flowing neutral clusters and is the only source of charge-dependent correlations, the $\Delta\gamma/v_{2}$ ratios with respect to different event planes ($\Psi_{A}$, $\Psi_{B}$, $\Psi_{C}\cdots$) are expected to be the same~\cite{Xu:2017qfs,Voloshin:2018qsm,Voloshin:2004vk,Schlichting:2010qia,Wang:2009kd,Schenke:2019ruo}, i.e.,
\begin{linenomath}
\begin{equation}
    \Delta\gamma/v_{2}(\Psi_{A})=\Delta\gamma/v_{2}(\Psi_{B})=    \Delta\gamma/v_{2}(\Psi_{C}) \cdots
\label{gammaratio:motivation}
\end{equation}
\end{linenomath}
For our measurements, in the case of a flow driven background scenario, one expects:
\begin{linenomath}
\begin{equation}
  \begin{split}
    \MoveEqLeft
    \Delta\gamma/v_{2}(\Psi_{1,\mathrm{Y_{beam}}<|\eta|<5.1}) \\
    &=\Delta\gamma/v_{2}(\Psi_{2,\,2.1<|\eta|<\mathrm{Y_{beam}}})=    \Delta\gamma/v_{2}(\Psi_{2,|\eta|<1}).
    \end{split}
    \label{eq_cmebkg}
\end{equation}
\end{linenomath}

The aim of our study is to test any deviation from the flow driven background scenario (Eq.~\ref{eq_cmebkg}) in Au+Au collisions at $\sqrt{s_{_{\rm NN}}}=27$ GeV. 
It has been argued that an observation of $\Delta\gamma/v_{2}(\Psi_1) >\Delta\gamma/v_{2}(\Psi_2)$  cannot be caused by flow driven  background~\cite{Xu:2017qfs,Voloshin:2018qsm}. This is because background from flowing resonances is largest along $\Psi_2$~\cite{Xu:2017qfs}. However, if $\Delta\gamma/v_{2}(\Psi_{1})$ is significantly larger than $\Delta\gamma/v_{2}(\Psi_{2})$, the observation would indicate larger magnetic field driven charge separation across $\Psi_1$ than that of $\Psi_2$. Such an observation will have an implication for the CME scenario since the CME signal is expected to be correlated with the magnetic field direction. It is important to note that Eq.~\ref{eq_cmebkg} is expected to be robust against event plane de-correlations or flow fluctuations, since they affect both numerator and denominator in the same way (see Ref.~\cite{Voloshin:2018qsm}). In addition, the effect of non-flow may cause deviations from Eq.~\ref{eq_cmebkg} as discussed in Ref.~\cite{Feng:2021pgf} which studied the effect using AMPT~\cite{Lin:2004en} and HIJING~\cite{Wang:1991hta} simulations. At the top RHIC energy, non-flow contamination to the CME-like signal due to sources such as fragmentation and momentum-conservation from dijets, is found to be $-5\pm3\%$ to $4\pm 5\%$, depending on the choice of the event planes. However, in the context of our analysis at lower energy, production of dijets is expected to be smaller and so non-flow effects are expected to be smaller at forward rapidity.

As a first step, we use the combination of the TPC with the inner EPDs ($|\eta|>\mathrm{Y_{beam}}$) to measure the charge separation across the $\Psi_1$ using the $\gamma$-correlator expressed in the scalar product method~\cite{Adler:2002pu} as: 
\begin{linenomath}
\begin{equation}
    \begin{split}
    \MoveEqLeft
    \gamma(\Psi_1)=\gamma^{\alpha, \beta}_{1,1,1,1} (\eta_{\alpha}, \eta_{\beta})(\Psi_{1,{\mathrm{Y_{beam}}<|\eta|<5.1}})\\
    &= \langle \mathrm{ \cos( \phi_{\alpha}(\eta_{\alpha}) + \phi_{\beta}(\eta_{\beta})} \\
    &\,\,\,\,\,\,\,\,\,\,\, - \mathrm{\Psi_{1,\mathrm{Y_{beam}}<\eta<5.1} - \Psi_{1,\mathrm{-Y_{beam}}>\eta>-5.1}  ) } \rangle \\
    &\equiv \frac{ \langle {Q_\mathrm{1,TPC}^{\alpha}Q_\mathrm{1,TPC}^{\beta}Q^{*}_\mathrm{1,EPDE}Q^{*}_\mathrm{1,EPDW} } \rangle }{\langle { Q_\mathrm{1,EPDE}Q^{*}_\mathrm{1,EPDW}}  \rangle }.
    \end{split}
\end{equation}
\end{linenomath}
Where the subscripts ``1,1,1,1" denote first order harmonics associated with the azimuthal angle of particles $\phi_\alpha$, $\phi_\beta$ from TPC ($|\eta|<1$), the event planes $\Psi_1$ from the inner EPD east and west, respectively. 
Here, we use the algebra based on $Q$-vectors~\cite{Bilandzic:2010jr}, defined as $Q_{\rm n}$ = $\sum^{M}\limits_{i=1} w_{i}$e$^{in\phi}/\sum^{M}\limits_{i=1} w_{i}$. The weight factor $w_{i}$ accounts for the imperfection in the detector acceptance in bins of $\eta-\phi$, $p_{T}$ (track-curvature), $V_z$, and centrality. $M$ refers to the number of particles in the analysis. When the particles ``$\alpha$" and ``$\beta$" are of same-sign and share the same acceptance $|\eta|<1$,  the $Q$-vector estimations require special treatment as follows. We estimate  
\begin{linenomath}
\begin{equation}
Q_{\rm 1,TPC}^{\alpha}Q_{\rm 1,TPC}^{\beta} = \frac{\left(\sum\limits_i w_i e^{i \phi_i}\right)^2 - \sum\limits_i w_i^2 e^{i2\phi_i}}{ \left(\sum\limits_i w_i\right)^2 - \sum\limits w_i^2}, 
\end{equation}
\end{linenomath}
where $Q^{\alpha}_{\rm n,TPC}(\eta_\alpha)$ and $Q^{\beta}_{\rm n,TPC}(\eta_\beta)$ denote charge dependent $Q$-vectors of particles at pseudorapidities $\eta_\alpha$ and $\eta_\beta$ within $|\eta|<$ 1
and for 0.2 $\leq$ $p_{T}$ $\leq$ 3.0 GeV/c using the TPC. Similarly, Q$_{n, \rm{EPDE/W}}$ refers to the $Q$-vectors obtained from the hits in the EPDs which require slightly different treatment. For the EPDs we use the number of MIPs corresponding to hits produced by particles as weights, and assume that they pass through the center of the tile. Since the sign of the directed flow changes inside the EPD acceptance, we need to weight the first order $Q$-vectors with a parameterization of the directed flow (sign-and-magnitude) as a function of pseudorapidity, approximated as $v_{1}(\Psi_1, \eta)=a_1\eta + a_3\eta^3$. Here $a_1$ and $a_3$, obtained by fitting data, are the linear are cubic parameters that capture the rapidity-odd nature of $v_1$. We also estimate the elliptic anisotropy of the particles at mid-rapidity with respect to the $\Psi_1$ plane from the EPDs  using the scalar product method as follows:
\begin{linenomath}
\begin{equation}
    \begin{split}
    \MoveEqLeft
    v_{2,1,1}(\Psi_{1}) \\
    &=\left<\cos\left(2\phi - \Psi_{1,Y_{\rm beam}<\eta<5.1} - \Psi_{1,\,-5.1<\eta<-Y_{\rm beam}}\right)\right>\\
    &\equiv\frac{\left<{Q_{\rm 2,TPC}Q^{*}_{\rm 1,EPDE}Q^{*}_{\rm 1,EPDW}}\right>}{\langle { Q_{\rm 1,EPDE}Q^{*}_{\rm 1,EPDE}}  \rangle }. 
    \end{split}
    \label{eq_v211}
\end{equation}
\end{linenomath}
Here the notation ``2,1,1" denotes the use of second order harmonic in front of the azimuthal angle of particle $\phi$ from mid-rapidity and first order harmonics associated with the $\Psi_1$ planes from the east and west EPDs.

As a second step, we follow a similar approach to measure charge separation with respect to the $\Psi_2$ of the produced particles dominated by forward participants:
\begin{linenomath}
\begin{equation}
    \begin{split}
    \MoveEqLeft
    \gamma(\Psi_{2,\,\, 2.1<|\eta| < \mathrm{Y_{beam}}})=\gamma^{\alpha, \beta}_{1,1,2} (\eta_{\alpha}, \eta_{\beta})(\Psi_{2,\,\, 2.1<|\eta| < \mathrm{Y_{beam}}})\\
    &=\langle \mathrm{ \cos( \phi_{\alpha}(\eta_{\alpha}) + \phi_{\beta}(\eta_{\beta}) - 2\Psi_{2,\,2.1<|\eta|<\mathrm{Y_{beam}}})} \rangle \\
    &\equiv\frac{ \langle {Q_{\rm 1,TPC}^{\alpha}Q_{\rm 1,TPC}^{\beta}Q^{*}_{\rm 2,EPDE} + Q_{\rm 1,TPC}^{\alpha}Q_{\rm 1,TPC}^{\beta}Q^{*}_{\rm 2,EPDW} } \rangle }{2 \sqrt{ \langle { Q_{\rm 2,EPDE}Q^{*}_{\rm 2,EPDW}} } \rangle }.
    \end{split}
\end{equation}
\end{linenomath}
Here the subscripts ``1,1,2" denote first order harmonics of the azimuthal angles of the particles ``$\alpha$", ``$\beta$" and second order harmonic associated with the event plane $\Psi_2$ from the outer EPD. We estimate the corresponding elliptic anisotropy coefficient relative to the $\Psi_2$ using the scalar product method as: 
\begin{linenomath}
\begin{equation}
    \begin{split}
    \MoveEqLeft
    v_{2}(\Psi_{2,\,\, 2.1<|\eta| < \mathrm{Y_{beam}}})\\
    &=\left<\cos\left(2\phi - 2\Psi_{2,\,2.1<|\eta|<Y_{\rm beam}}\right)\right> \\
    &\equiv\frac{\langle {Q_{\rm 2,TPC}Q^{*}_{\rm 2,EPDE} + Q_{\rm 2,TPC}Q^{*}_{\rm 2,EPDW} } \rangle}{2 \sqrt{ \langle { Q_{\rm 2,EPDE}Q^{*}_{\rm 2,EPDW}} } \rangle }.
    \end{split}
    \label{eq_v2sp}
\end{equation}
\end{linenomath}

As a third step, we perform another measurement using charge separation across the elliptic flow plane of produced particles at mid-rapidity $|\eta|<1$ in the following way:
\begin{linenomath}
\begin{equation}
    \begin{split}
        \MoveEqLeft
    \gamma(\Psi_{2,|\eta|<1})=\gamma^{\alpha, \beta}_{1,1,2}(\eta_{\alpha} , \eta_{\beta}) (\Psi_{2,|\eta|<1}) \\
    &= \frac{\langle \cos(\phi_{\alpha}(\eta_{\alpha}) + \phi_{\beta}(\eta_{\beta}) - 2\phi_{c})\rangle }{v_{2,c}\{2\}} \\
    &\equiv \langle { \cos( \phi_{\alpha}(\eta_{\alpha}) + \phi_{\beta}(\eta_{\beta}) - 2\Psi_{2,|\eta|<1}) } \rangle.
    \end{split}
\end{equation}
\end{linenomath}
Similar to previously used convention, here the  subscripts ``1,1,2" associated with the notation of $\gamma$ refer to the order of harmonics in front of the azimuthal angles $\phi$ of three distinctly different particles ``$\alpha$, $\beta$", and ``$c$", all measured by the TPC. We use the charge-inclusive reference particle ``$c$" to construct the elliptic flow plane $\Psi_{2}$ at mid-rapidity. The quantity $v_{2,c}\{2\}$ (written as $v_2\{2\}$ in the following) refers to the elliptic flow coefficient of the reference particle `c' that we estimate using two-particle correlations:
\begin{linenomath}
\begin{equation}
    v_{2}\{2\}^2 (|\eta|<1)=\left<\cos(2\phi_1 (\eta_1)-2\phi_2(\eta_2))\right>.
    \label{eq_v2pc}
\end{equation}
\end{linenomath}
In this $v_{2}\{2\}^2$ measurement from the TPC, we require $\Delta\eta_{1,2}=|\eta_1-\eta_2|>0.05$ to remove track merging and electron pairs from photon conversions. For $v_{2}\{2\}^2(\Delta\eta_{1,2})$ measurements we also remove short-range components due to femtoscopic correlations using the approach described in Ref.~\cite{STAR:2016vqt}.

We perform measurements of $\gamma$ in $\Delta\eta=\eta_\alpha-\eta_\beta$ slices and require $\Delta\eta>0.05$ before integrating over $\Delta\eta$ to correct for the two-track merging effects which is most dominant in central collisions. The main challenge, when all three particles ``$\alpha,\beta$" and ``c" are taken from the TPC, is that no $\Delta\eta$ cut is applied between $\alpha,\beta$ and ``c" to reduce short-range non-flow correlations. This can be circumvented using a sub-event method by restricting, for example, ``c" from $-1<\eta<-0.2$ and ``$\alpha,\beta$" from $0.2<\eta<1$. However, restricting the acceptance of ``$\alpha$, $\beta$" results in larger statistical uncertainty that is particularly problematic at 27 GeV due to the lower number of produced particles compared with higher collision energies. We therefore avoid using the sub-events method. This difficulty highlights the advantage of using event planes from the EPDs at low energy which helps suppress short-range correlations while using the full TPC acceptance for $\alpha$ and $\beta$ to get the highest statistical significance.

In our measurements, we determine centrality using the probability distributions of uncorrected TPC tracks within $|\eta| < $ 0.5. We use a two-component Monte Carlo Glauber model fit to determine the values of average number of participating nucleons N$_{\rm{part}}$ in nine centrality intervals (0-5\%, 5-10\%, 10-20\%, ... , 70-80\%). Scaling the correlation observables by the number of participants N$_{\rm part}$ as written in Eq.~\ref{fig:gamma_ratio_byv2} compensates for the natural dilution of correlations ($\Delta\gamma \sim 1/N_{\rm part}$) due to an increasing number of  superposition of independent sources while going from peripheral to central events~\cite{Trainor:2000dm}. 

\section{Statistical and Systematic uncertainties}
\label{sec:syserr}
We use standard error propagation method for statistical uncertainty estimations in our analysis. However, for ratio observables such as $\Delta\gamma/v_2$ we examine the contribution from covariance terms. For this, we use an analytical approach as well as a Monte Carlo approach that is equivalent to the statistical Bootstrap method~\cite{10.1214/aos/1176344552}, originally developed for the STAR isobar blind analysis~\cite{STAR:2021mii}.  Analytical estimates indicate that the statistical uncertainty in the quantity $\Delta\gamma/v_2$ is dominated by the numerator (a factor of 50 larger than the co-variance term) and the co-variance terms can be ignored~\cite{STAR:2021mii}. Monte Carlo approach also leads to a consistent conclusion. The statistical uncertainties for all the results presented in this letter are obtained using the method of error propagation. Our study of the ratio of $\Delta\gamma/v_2$ with respect to $\Psi_1$ and $\Psi_2$ planes shows that the analytical method of error propagation ignoring co-variance overestimates the statistical uncertainty by $5\%$ using two different Monte Carlo methods in $10-50\%$ centrality (see supplementary material). 

The systematic uncertainties in our measurements include contributions from different choices of track and event selection conditions. We use the Barlow method to remove the effects of statistical fluctuations in the systematic error estimation~\cite{Barlow:2002yb}. For details of the Barlow method, see Ref.~\cite{STAR:2021mii}. The relative uncertainty number quoted for each case for the purpose of the  following discussion are estimated for the final observable of interest that is the double-ratio of ($\Delta\gamma/v_2$) with respect to $\Psi_1$ and $\Psi_2$ planes within 10-20$\%$, 20-30$\%$, 30-40$\%$, and 40-50$\%$ centrality bins. A variation of the minimum number of ionization points in the TPC from $15$ to $20$ leads to a relative systematic uncertainty up to $7\%$. We find that a variation of the global DCA of the track to the primary vertex from $<3$ cm to $<2$ cm leads to a contribution up to $1\%$. Systematic errors arise due to trigger bias and changes in beam luminosity. This we estimate by separately analyzing low, middle and high luminosity data sets and find a contribution up to $0.2\%$. The uncertainty associated with the determination of EPD $\Psi_{1}$ is obtained by varying the acceptance from the default cut of 3.4 (Y$_{\rm beam}) <\eta<5.1$ to $4.0<\eta<5.1$, which leads to a contribution up to $2\%$. The variation from 3.4 (Y$_{\rm beam}) <\eta<5.1$ to full EPD acceptance $2.1<\eta<5.1$ leads to a contribution up to $0.5\%$ for the systematics associated with the $\Psi_{1}$ plane estimation. 
Similarly, we vary the acceptance for determining the $\Psi_{2}$ plane from the default cut of $2.1<\eta< 3.4$ (Y$_{\rm beam}$) to $2.1<\eta<3.0$ and $2.1<\eta<5.1$ (full EPD) leading to systematic uncertainties of $2\%$ and $0.5\%$, respectively. 
 We add different systematic uncertainty sources in quadrature and obtain the total systematic uncertainty is not bigger than $7\%$. 
\begin{figure}[t]
    \centering
   \includegraphics[width=0.46\textwidth]{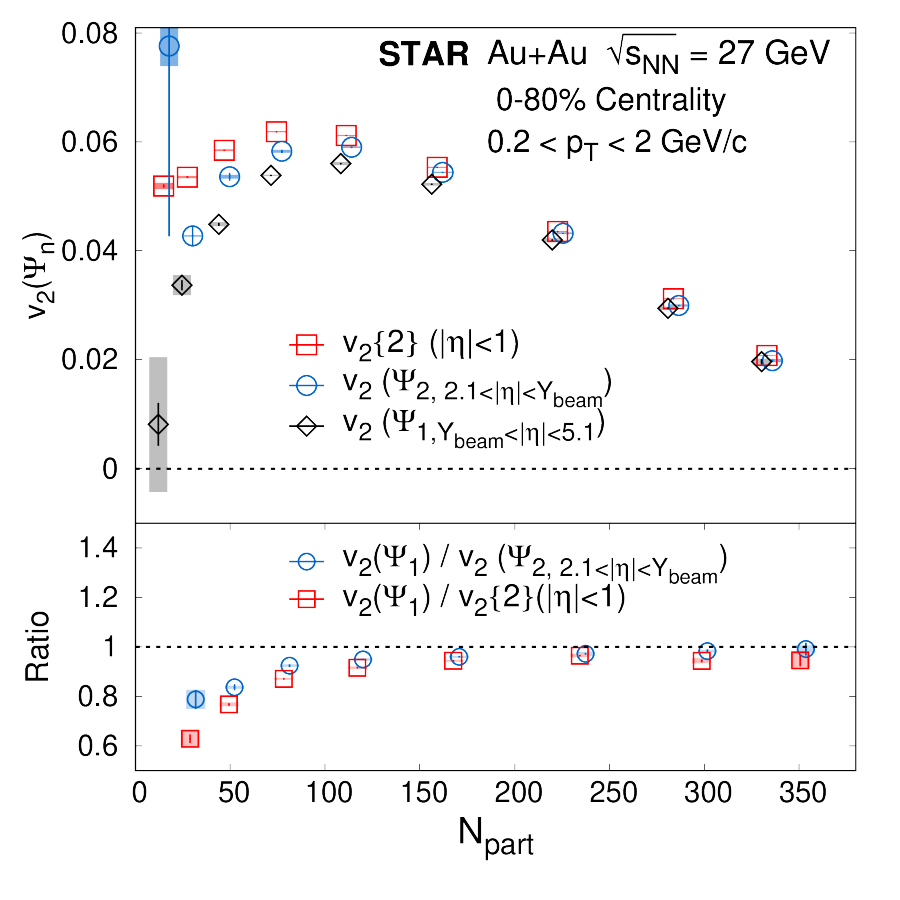}
    \caption{Elliptic anisotropy coefficient $v_{2}$ using TPC tracks and EPD hits. The upper panel shows elliptic flow relative to different event planes. The lower panel shows the $v_{2}$ ratio using the directed flow plane compared with the forward and mid rapidity elliptic flow planes. The lines indicate the statistical uncertainties and the shadowed boxes indicate the systematic uncertainties. The centrality bins are shifted horizontally for clarity.
    }
    \label{fig:v2_ratio}
\end{figure}

\begin{figure}[htb]
    \centering
    \includegraphics[width=0.5\textwidth]{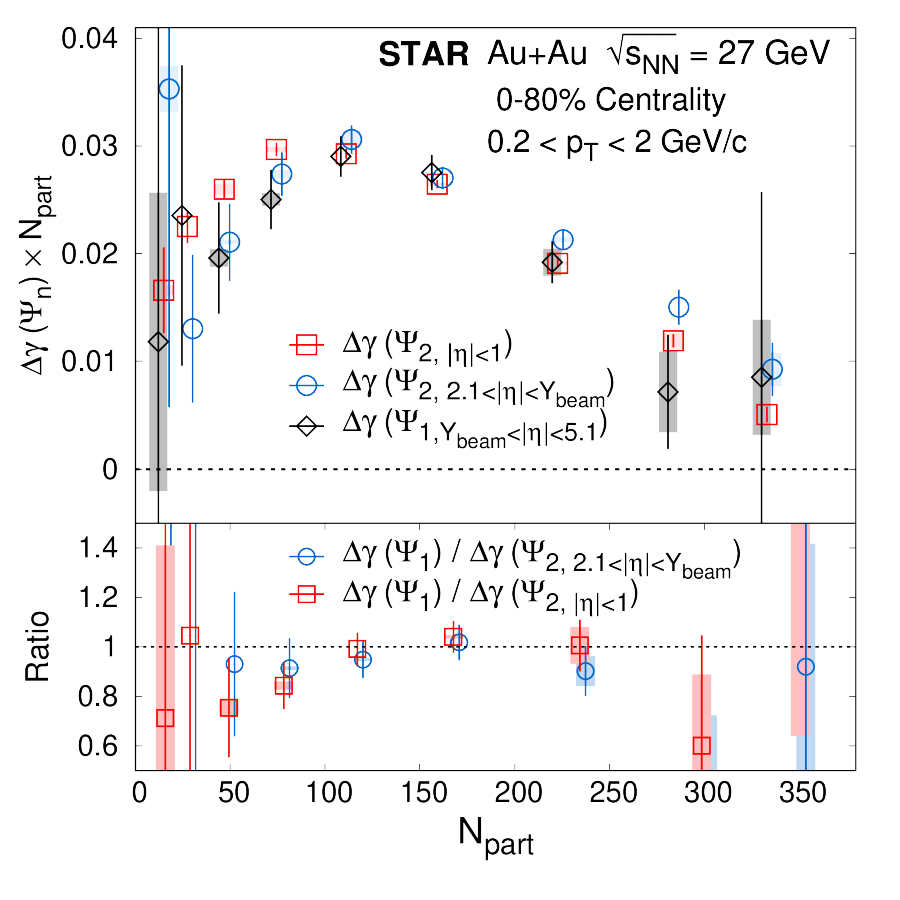}
    \caption{Charge separation across different event planes measured by the difference between opposite (OS) and same sign (SS) $\gamma$-correlators. The upper panel shows the $\Delta\gamma=\gamma$ (OS) - $\gamma$ (SS) across different event planes. The $\Delta\gamma$ points are scaled with N$_{\rm part}$ to account for dilution due to super-position of independent sources and also to improve the visibility. The lower panel shows the ratios of $\Delta\gamma$ across the directed flow plane to the same across the elliptic flow planes. 
    The lines indicate the statistical uncertainty, the shadowed boxes  indicate the systematic uncertainty. Points are shifted horizontally for clarity.}
    \label{fig:gamma_ratio}
\end{figure}
Similar to previous STAR analyses of three-particle correlations~\cite{STAR:2021pwb,STAR:2021mii, STAR:2017idk,Adamczyk:2017byf}, we investigate the effects of the following cut variations: 1) the acceptance of tracks from the default value of $|\eta|<1.0$ to $|\eta|<0.8$, 2) variation of the V$_{z}$ cut from the default value of $|V_{z}| <$ 40 cm to 0 $<$ V$_{z}$ $<$ 40 cm and, 3) variation of the maximum allowed V$_{r}$ from 2 cm to 1 cm. In all such cases, we find zero systematic uncertainty following the Barlow criteria. In addition, we study the effect of $p_T$ dependent tracking efficiency by using it as a weight ($w_i$) for the $Q$-vectors from the TPC. We observe no statistically significant difference in our final observables with and without efficiency weights. For the results shown in this letter we do not include $p_T$ dependent tracking efficiency as weights. We studied the stability of the results by changing the threshold on the number of MIPs for EPD hits in the $Q$-vector estimation. After changing the value of MIP threshold from 0.3 to 1, we do not see any statistically significant change in our results.

\section{Results}
\label{sec:result}
In Fig.~\ref{fig:v2_ratio}, we show the elliptic flow coefficients \\
$v_{2,1,1}(\Psi_{1, |\eta|>Y_{\rm beam}})$, $v_{2}(\Psi_{2, |\eta|<Y_{\rm beam}})$, and $v_{2}(\Psi_{2,|\eta|<1})$ defined in Eq.~\ref{eq_v2sp}-\ref{eq_v2pc} as a function of $N_{\rm part}$ in the upper panel, and the ratios of $v_2(\Psi_1)/v_2(\Psi_2)$ in the lower panel. The difference in the magnitudes of $v_2$ from mid-rapidity to forward rapidity can be attributed to changes in the non-flow contribution, flow fluctuations, and event plane de-correlation. It is challenging to disentangle these three effects as was discussed in previous STAR publications such as Ref.~\cite{STAR:2011ert,STAR:2014ofx}. The lower panel of Fig.~\ref{fig:v2_ratio} indicates a drop of $20-40\%$ in $v_{2}$ along the $\Psi_1$ plane in comparison to the same from the $\Psi_2$ plane for peripheral events.

In Fig.~\ref{fig:gamma_ratio}, we show the charge-dependent $\gamma$ correlator, $\Delta\gamma$ = $\gamma_{OS} - \gamma_{SS}$,
measured relative to $\Psi_{2,|\eta|<1}$ plane,  $\Psi_{2,2.1<|\eta|<Y_{\rm beam}}$ plane, and  $\Psi_{1,Y_{\rm beam}<|\eta|<5.1}$ as a function of $N_{\rm part}$.
In mid-central events the magnitudes of $\Delta\gamma$ for different planes are consistent with each other. In central and peripheral events, results for the $\Psi_1$ plane hint at a weaker charge separation although differences are smaller than the statistical uncertainties. This is also evident from the ratio plot shown in the lower panel. It is difficult to make any conclusion related to the magnetic field driven charge separation from $\Delta\gamma$ ratio, as a flow-driven background is the dominant contribution to the $\Delta\gamma$ correlator. 
\begin{figure}[t]
    \centering
    \includegraphics[width=0.5\textwidth]{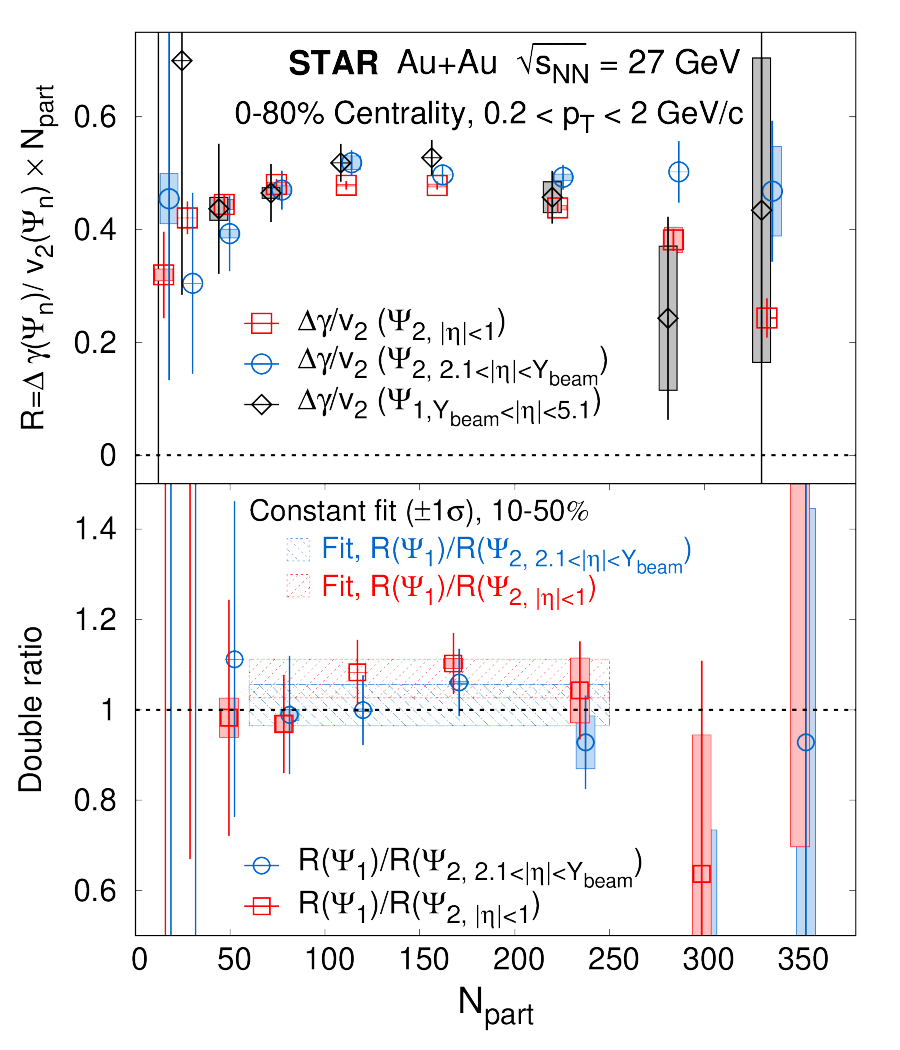}
    \caption{(Upper panel) The quantity $R$ obtained by estimating the charge separation measured by the difference between opposite-sign and same-sign $\gamma$ correlator ($\Delta\gamma$) and then scaling by $v_2$ times $N_{\rm part}$. The measurements are shown for all three different event planes. (Lower panel) The ratio of the quantity $R$ shown on upper panel between $\Psi_1$ plane and $\Psi_2$ plane. The error bars indicate the statistical uncertainty, the shadowed bars indicate the systematic uncertainty. The results of the fit including $1\sigma$ fitting uncertainties are shown by bands with dashed border for $10-50\%$ centrality in lower panel. Points are shifted horizontally for clarity.}
    \label{fig:gamma_ratio_byv2}
\end{figure}

Therefore, in Fig.~\ref{fig:gamma_ratio_byv2}, upper panel, we present the centrality dependence of the quantity
\begin{linenomath}
\begin{equation}
R(\Psi_n)=\frac{\Delta\gamma(\Psi_n)}{v_{2}(\Psi_n)}\times N_{\rm part},
\end{equation}
\end{linenomath}
measured relative to the $\Psi_1$ and $\Psi_2$ planes from forward and mid-rapidity. Compared to the $\Delta\gamma$ measurements shown in Fig.~\ref{fig:gamma_ratio}, we observe a much weaker centrality dependence after scaling $\Delta\gamma$ with $v_2$.

Finally, to quantify the difference between the charge separation across $\Psi_1$ plane
relative to that of $\Psi_2$, we take a ratio between $R(\Psi_1)$ and $R(\Psi_2)$. As mentioned in  Eq.~\ref{eq_cmebkg}, $R(\Psi_1)/R(\Psi_2)$ is expected to be unity in the case of flow driven background scenario. For magnetic field driven correlations, we expect this ratio to be above unity.
The values of $R(\Psi_1)/R(\Psi_2)$ are shown on the lower panel of Fig.~\ref{fig:gamma_ratio_byv2}. We fit this quantity over a centrality range of $10-50\%$ using a constant function by properly incorporating the statistical and systematic uncertainties. 

To quantitatively estimate the deviation from a flow driven background, we define a quantity $\mathcal{D}$ as follows,
\begin{linenomath}
\begin{equation} 
\mathcal{D}=R(\Psi_1)/R(\Psi_2)-1,    
\end{equation}
\end{linenomath}
where observation of a significant nonzero value of $\mathcal{D}$ implies the presence of the magnetic field driven correlations~\footnote{ 
According to previous studies~\cite{Xu:2017qfs, Voloshin:2018qsm} one can obtain a relation like:
\begin{equation}
           \mathcal{D}=
           \frac{(\Delta\gamma/v_{2})_{\Psi_1}}{(\Delta\gamma/v_{2})_{\Psi_2}} - 1 = f_{\rm CME}({\Psi_2}) \left(\frac{\gamma_{B}({\Psi_1})}{\gamma_{B}({\Psi_2})}\frac{v_{2}({\Psi_2})}{v_{2}({\Psi_1})}-1 \right).
           \label{D_fcme}
\end{equation}
In the flow-driven background scenario, one expects $f_{\rm CME}=0$, therefore $\mathcal{D}$ is expected to be zero. However, in the presence of CME one expects $\mathcal{D}>0$. This is because the elliptic flow is maximum w.r.to $\Psi_2$ plane. So, we always have $v_{2}(\Psi_2)/v_{2}(\Psi_1)>1$ (measurement).  Also, since $\Psi_1$ is determined by the directed flow of forward protons (which also generate B-field), it has a larger correlation with the B-field direction than the $\Psi_2$ plane. Therefore, $\gamma_{B}({\Psi_1})/\gamma_{B}({\Psi_2})>1$ (UrQMD simulation). In the presence of CME one has $f_{\rm CME}>0$ so $\mathcal{D}>0$.

}.
\begin{figure}[t]
\vspace{20pt}
    \centering
    \includegraphics[width=0.5\textwidth]{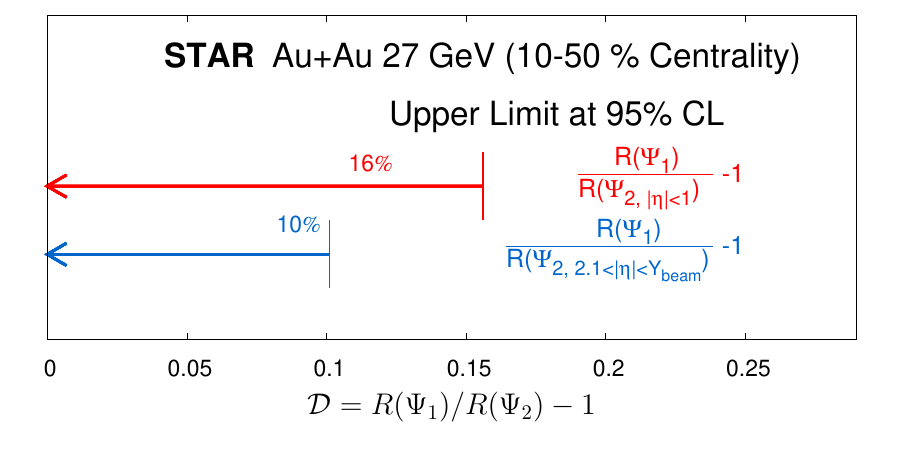}
    \caption{
    The upper limit at the $95\%$ CL calculated for the deviation quantity $\mathcal{D}$ in 10-50\% centrality.
    }
    \label{fig:confidence_level}
\end{figure}
We find the quantity $\mathcal{D}$ to be $0.011\pm0.046$ and $0.069\pm0.043$ when $R(\Psi_2)$ is estimated from $\Psi_{2,2.1<|\eta|<Y_{\rm beam}}$ and  $\Psi_{2,|\eta|<1}$, respectively. The results are consistent with zero within total uncertainty. In order to quantify the possible deviation from zero, we derive an upper limit at the $95\%$ CL on $\mathcal{D}$ using the { Feldman-Cousin approach}~\footnote{ The upper limit of a quantity with measured mean $x_0$ and standard deviation $\sigma$ at 95\% confidence is: $x_0\, + \, (1.96 + n(x_0))\times \sigma$. The quantity $n(x_0)$ can be obtained from table.10 of Ref~\cite{Feldman:1997qc}.} discussed in Ref.~\cite{Feldman:1997qc}. In centrality 10-50$\%$, we find the upper limits of $\mathcal{D}$ to be 10$\%$ and 16$\%$ using $\Psi_2$ at forward and mid-rapidity, respectively. Here the numbers are rounded into integers. We show these estimates in Fig.~\ref{fig:confidence_level}.

\section{Summary}
\label{sec:summary}
In this letter, we present measurements of charge separation with respect to reaction plane using Au+Au $\sqrt{s_{_{\rm NN}}}$ $=$27 GeV collisions.  At this collision energy, the STAR forward EPDs can measure the event plane associated with a large directed flow from beam fragments with high resolution. This directed flow plane ($\Psi_1$) is expected to be more correlated with the direction of magnetic field than the elliptic flow planes ($\Psi_2$) of produced particles as shown in UrQMD simulations. Therefore, we measured the charge separation scaled by ellipticity $R(\Psi_n)$ across the $\Psi_1$ determined at forward rapidity and compare it with the corresponding measurements using $\Psi_2$ reconstructed out of produced particles at both forward and mid-rapidity.  
Within our measurement uncertainties we find the ratio $R(\Psi_1)/R(\Psi_2)$ to be consistent with unity, which agrees with the expectations from a flow driven background scenario. For further quantification, we derive an upper limit at the 95\% confidence level on the quantity $\mathcal{D}=R(\Psi_1)/R(\Psi_2)-1$ for
possible deviation from a flow driven background scenario. In 10-50$\%$ centrality we find the upper limits of $\mathcal{D}$  to be 10$\%$ and 16$\%$ using $\Psi_2$ at forward and mid-rapidity, respectively. 
In this analysis we argued that 
the information of the directed flow near the beam rapidity using EPDs and the elliptic flow at mid-rapidity allows us to control the flow driven CME background in 27 GeV Au+Au collisions and explore effects beyond background. Due to the acceptance of the EPDs ($2.1<\eta<5.1$), the same can be done for several data sets ($\sqrt{s_{_{NN}}}=9.2, 11.5, 13.7, 14.5, 17.3, 19.6$ GeV) collected by the STAR under RHIC Beam Energy Scan Phase II program. 
The use of forward event planes in this work not only pioneers a high-precision CME search from the RHIC Beam Energy Scan Phase II program, but also opens up opportunities to search for other magnetic field driven effects at RHIC.

\section*{Acknowledgement:}
We thank the RHIC Operations Group and RCF at BNL, the NERSC Center at LBNL, and the Open Science Grid consortium for providing resources and support.  This work was supported in part by the Office of Nuclear Physics within the U.S. DOE Office of Science, the U.S. National Science Foundation, National Natural Science Foundation of China, Chinese Academy of Science, the Ministry of Science and Technology of China and the Chinese Ministry of Education, the Higher Education Sprout Project by Ministry of Education at NCKU, the National Research Foundation of Korea, Czech Science Foundation and Ministry of Education, Youth and Sports of the Czech Republic, Hungarian National Research, Development and Innovation Office, New National Excellency Programme of the Hungarian Ministry of Human Capacities, Department of Atomic Energy and Department of Science and Technology of the Government of India, the National Science Centre and WUT ID-UB of Poland, the Ministry of Science, Education and Sports of the Republic of Croatia, German Bundesministerium f\"ur Bildung, Wissenschaft, Forschung and Technologie (BMBF), Helmholtz Association, Ministry of Education, Culture, Sports, Science, and Technology (MEXT) and Japan Society for the Promotion of Science (JSPS).

\bibliographystyle{elsarticle-num-names}
\bibliography{elsarticle-main}

\newpage
\onecolumn
\section*{ Supplementary Material: a novel fast Monte-Carlo method to estimate the statistical uncertainty of a ratio observable used in this letter}

In this measurement, a data driven Monte Carlo (MC) method has been used to quantify the correlated uncertainties in the ratio quantity $\rm I\!R=\frac{\rm R(\Psi_1)}{\rm R(\Psi_2)}$, where ${\rm R}=\Delta\gamma/v_2=\langle\cos{(\phi^{\alpha}+\phi^{\beta}-2\Psi)}\rangle/\langle\cos(\phi-\Psi)\rangle$ as described in the main text. In this quantity, although the event planes ($\rm \Psi$) are estimated with different acceptance in $\rm R(\Psi_1)$ and $\rm R(\Psi_2)$, the particles of interest for $\Delta\gamma$ ($\phi_{\alpha}$, $\phi_{\beta}$) and $v_{2}$ ($\phi$) measurements are from the same TPC acceptance. Thus the possible anti-correlation/correlation in the variance of the ratio needs to be examined. 

In high energy physics, to study the statistical uncertainties, the most widely used Monte Carlo method is called the ``Bootstrap method"~\cite{Efron:bootstrap}. In this study, we designed a new method specific for ratio quantities $\left<x\right>/\left<y\right>$. We call this new approach as the ``AB method" which is computationally economical. We have also checked the consistency of our approach with the classical Bootstrap method using experimental data. To perform this consistency check we have used about one third of the whole statistics. 

For the Bootstrap method we follow the approach described in Ref~\cite{Efron:bootstrap}. The Bootstrap approach requires creating copies of the data sample through Monte Carlo sampling in which some of the events will be duplicated while some will be absent, by construction. We perform this sampling procedure $N$ times to get a distribution of the ratio observable $P^{Bootstrap}(\rm I\!R)$. From the distribution we estimate the mean $\mu^{Bootstrap}$ and width $\sigma^{Bootstrap}$ of the ratio ${\rm I\!R}$.

For the AB method, we divide the entire data sample into two halves. We call the two halves ``group A" and ``group B". For the ratio $\rm I\!R= R(\Psi_1)/R(\Psi_2)=\left<x\right>/\left<y\right>$, we estimate $x$ and $y$ from the two groups and label them as $\left<x(A)\right>$, $\left<x(B)\right>$, $\left<y(A)\right>$, and $\left<y(B)\right>$. Thus we can estimate the ratios  $\left<x(A)\right>/\left<y(A)\right>$, $\left<x(A)\right>/\left<y(B)\right>$, $\left<x(B)\right>/\left<y(A)\right>$, and $\left<x(B)\right>/\left<y(B)\right>$. When the $\left<x\right>$ and $\left<y\right>$ come from the same half ($\left<x(A)\right>/\left<y(A)\right>$ and $\left<x(B)\right>/\left<y(B)\right>$), we call the ratios ``AB-same" and when they are came from the different halves ($\left<x(A)\right>/\left<y(B)\right>$ and $\left<x(B)\right>/\left<y(A)\right>$), we call them ``AB-cross". Note, in this case each sample gives us two entries for both AB-same and AB-cross. We repeat the sampling procedure $N$ times to get the probability distributions for AB-same ($P^{AB-same}({\rm I\!R})$) and AB-cross ($P^{AB-cross}(\rm I\!R$). For the AB-same we can estimate the mean $\mu^{AB-same}$ and width $\sigma^{AB-same}$. Similarly, for AB-cross we can estimate the mean $\mu^{AB-cross}$ and the width $\sigma^{AB-cross}$. 

The Bootstrap method is expected to lead to a variance of ratio similar to the analytical expression of variance ($\sigma^2$) including correlated fluctuations: \\
 \begin{equation}
\begin{split}
&\sigma^{Bootstrap} \approx \, (\sigma_{x}^2 (\partial{\rm I\!R}/\partial{\left<y\right>})^2 + \sigma_{y}^2 (\partial{\rm I\!R}/\partial{\left<x\right>})^2\\
&+ 2 \rho \sigma_{y} \sigma_{x}  (\partial^{2}{R}/\partial{\left<y\right>}\partial{\left<x\right>}))^{\frac{1}{2}}, \\ 
\end{split}
\label{eq_bootstrap}
     \end{equation}
where $\sigma_{x}$ and $\sigma_{y}$ are the widths of the distributions of the numerator and denominator, respectively. $\rho$ is the correlation coefficient. The $\sigma^{AB-same}$ should have the expression as Eq.~\ref{eq_bootstrap}.
For AB-cross sample, it should be: 
 \begin{equation}
    \sigma^{AB-cross} = \, (\sigma_{x}^2 (\partial{\rm I\!R}/\partial{\left<y\right>})^2 + \sigma_{y}^2 (\partial{\rm I\!R}/\partial{\left<x\right>})^2 ))^{\frac{1}{2}}, \\      
\label{eq_abcross}
\end{equation}
there is no correlation term in contrast to Eq.~\ref{eq_bootstrap} because these samples of $x$ and $y$ are uncorrelated. 

Our expectations are the following:
    \begin{enumerate}
        \item All the three cases should give rise to the same value of mean. \\
        \begin{equation}
                \mu^{AB-same} = \mu^{AB-cross}= \mu^{Bootstrap} 
        \label{exp-1}   
        \end{equation}
 
        \item If there is an anti-correlation, we should get: \\
        \begin{equation}
            \sigma^{AB-same} \approx \sigma^{Bootstrap} > \sigma^{AB-cross}
            \label{exp-2}
        \end{equation}
        
        \item If there is a correlation, we should get: \\
        \begin{equation}
            \sigma^{AB-cross} > \sigma^{Bootstrap} \approx \sigma^{AB-same}
        \label{exp-3}
        \end{equation}
    \end{enumerate}    
The expectations of Eq.~\ref{exp-2},\ref{exp-3} for $\sigma^{AB-cross}$ are very easy to understand. Since  $\sigma^{AB-cross}$ is estimated from two independent data sets, there should be no co-variance between the numerator and the denominator. Therefore, in the presence of correlations ($\rho>0$) and anti-correlations ($\rho<0$), the variance of ratio of the terms from two independent data sets will be over and underestimated, respectively. The expectations that $\sigma^{AB-same}$ and $\sigma^{Bootstrap}$ are approximately equal is not straightforward but can be easily demonstrated by Monte Carlo simulations as follows.

\begin{figure}[htb]
    \centering
    \includegraphics[width=0.4\textwidth]{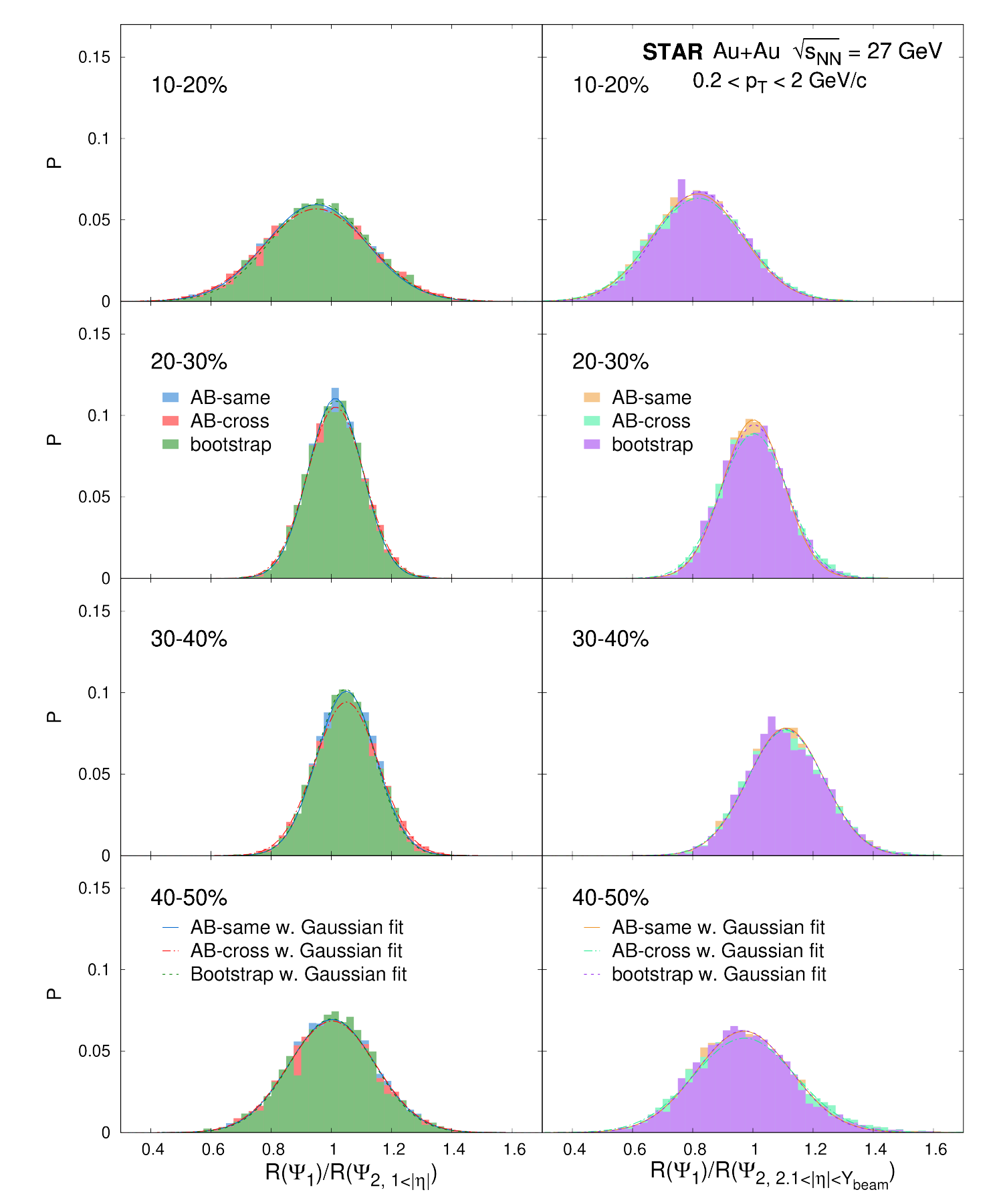}
    \caption{The distribution for the AB-same, AB-cross, and Bootstrap method after 3000 times sampling for $10-50\%$ centrality. The histograms are fitted with Gaussian distributions as shown by lines with different colors. All the distributions have a similar mean. The AB-same and Bootstrap distributions correspond to the correct variance. The wider distributions for AB-cross include correlations.}
    \label{fig1:gaussian_hist}
\end{figure}

The results from our exercise are shown in Fig.~\ref{fig1:gaussian_hist} in terms of the distributions of the ratio observable $P(\rm I\!R$) in different centralities and acceptance after sampling 3000 times. The left side panels are for the measurements on ${\rm R(\Psi_1)}/{\rm R(\Psi_{2,~1<|\eta|})}$, the right side panels are the measurements on ${\rm R(\Psi_1)}/{\rm R(\Psi_{2,~2.1<|\eta|<{\rm Y_{beam}}})}$. The histograms are fitted with Gaussian distributions. The AB-same and Bootstrap give very similar results as expected (see Eq.~\ref{exp-2} and \ref{exp-3}). The relative differences between the widths obtained from these two methods are consistent within $1\%$. From our exercise we observe a slightly wider width for the AB-cross case, which indicates the presence of correlated fluctuations as per Eq~\ref{exp-3}. From the AB-cross results, we find the width difference is less than $5\%$ compared to the AB-same case in $10-50\%$ centrality. 

The width of the distribution is proportional to the statistical uncertainty in the measurements of the ratio. We have established the consistency between AB-same and Bootstrap. Therefore, according to Eq.\ref{eq_bootstrap} and Eq.\ref{eq_abcross} the difference in the widths between AB-same and AB-cross method is an estimate between the true statistical uncertainty and the one ignoring the co-variance term in error propagation. Our exercise indicate the presence of correlated fluctuations and that as a result, the analytical method of error propagation ignoring co-variance overestimates the statistical uncertainty in the quantity $\rm I\!R$ by $5\%$. 

\end{document}